# Bound-extended mode transition in type-II synthetic photonic Weyl heterostructures

Wange Song[1], Zhiyuan Lin[1], Jitao Ji[1], Jiacheng Sun[1], Chen Chen[1], Shengjie Wu[1], Chunyu Huang[1], Luqi Yuan[2*], Shining Zhu[1], and Tao Li[1*]

[1]National Laboratory of Solid State Microstructures, Key Laboratory of Intelligent Optical Sensing and Manipulation, Jiangsu Key Laboratory of Artificial Functional Materials, College of Engineering and Applied Sciences, Nanjing University, Nanjing 210093, China.

[2]State Key Laboratory of Advanced Optical Communication Systems and Networks, School of Physics and Astronomy, Shanghai Jiao Tong University, Shanghai 200240, China.

*Corresponding authors: *yuanluqi@sjtu.edu.cn*; *taoli@nju.edu.cn*

## Abstract:

Photonic structures with Weyl points (WPs), including type-I and type-II, promise nontrivial surface modes and intriguing light manipulations for their three-dimensional topological bands. While previous studies mainly focus on exploring WPs in a uniform Weyl structure, here we establish Weyl heterostructures (i.e., a nonuniform Weyl lattice) with different rotational orientations in the synthetic dimension by nanostructured photonic waveguides. In this work, we unveil a transition between bound and extended modes on the interface of type-II Weyl heterostructures by tuning their rotational phases, despite the reversed topological order across the interface. This mode transition is also manifested from the total transmission to total reflection at the interface. All of these unconventional effects are attributed to the tilted dispersion of type-II Weyl band structure that can lead to mismatched bands and gaps across the interface. As a comparison, the type-I Weyl heterostructures lack the phase transition due to the untilted band structure. This work establishes a flexible scheme of artificial Weyl heterostructures that opens a new avenue towards high-dimensional topological effects and significantly enhances our capabilities in on-chip light manipulations.

The Weyl points (WPs) are degeneracy points of three-dimensional (3D) band structure with linear dispersion in the momentum space. They act as sources or drains of the Berry curvature and imply the emergence of high-dimensional topological modes known as the Fermi arc surface states [1-15]. As a fundamental topological phenomenon, WPs and associated surface states have been demonstrated in various physical systems, such as electronics [2,3], acoustics [4-7], and photonics [8-15]. In most of the previous studies, the Fermi arc states are observed at the surface of a uniform Weyl medium. Recently, new interface modes between two independent Weyl structures have been demonstrated in photonic lattices [16,17] with the assistance of synthetic dimension [18-25], which bring about new insights and explorations for richer interface physics of Weyl lattices.

Depending on the shape of the Fermi surface (isofrequency surface in the photonics context) at the Weyl frequencies, Weyl systems can be categorized as type I with a point-like Fermi surface and type II with a conical one [12,13,26-29]. The dramatic difference in band structure between the type-I and type-II Weyl media has brought abrupt physics consequences [30]. In the viewpoint of topological feature, robust edge states that are widely presented at the boundaries or interfaces attribute to the band inversion across two different media, which has been revealed in two type-I Weyl media by reversing their topological orders [16,17]. In contrast, the type-II Weyl media with a strongly anisotropic dispersion imply new possibilities by constructing the Weyl heterostructures in the synthetic space, in which the topological edge state at the interface remains unidentified and even its existence needs to be explored. The dispersive Weyl band structure incorporation with the reversed topological orders would possibly set new rules for the emergence of topological modes and inspire new physics and effects. Moreover, the experimental realization of a type-II Weyl heterostructure in photonic systems still remains a big challenge.

In this work, we reveal a bound-extended mode transition and showcase total reflection phenomena in type-II Weyl heterostructure, which are constructed by synthetic dimension in nanostructured silicon-on-insulator (SOI) waveguides [31,32]. For the type-II Weyl heterostructures consisting of two type-II WPs with opposite topological charges, the topological modes would disappear in some parameter conditions regardless of the reversed topological order, which arises from the tilted dispersion of type-II Weyl band structure. This is in contrast to the type-I cases, where the topological modes always exist as long as the topological order is reversed. A 2D phase diagram is depicted to show the phase transition of bound-extended modes. We further show that the tilted dispersion can lead to total reflection at the type-II interface, differing from the type-I case with transmission across the interface. All of these results are observed in SOI experiments at telecommunication wavelengths entirely consistent with theoretical predictions.

We start with explaining the schematics of the synthetic Weyl lattice in the integrated photonic platform [33-45] with etched longitudinal subwavelength grating (SWG) silicon waveguides (Fig. 1(a)), where $a$ is the width of the silicon slab, $P$ is the grating period. The cross-section of a unit cell is shown in Fig. 1(a), which consists of two waveguides with the widths defined as $w_1=w_{1c}(1+f_1l)$, $w_2=w_{2c}(1-f_2l)$, the gaps $d_1=d_c(1+m)$, $d_2=d_c(1-m)$, and the lattice constant $\Lambda=w_{1c}+w_{2c}+2d_c$. Here, $d_c=0.45$ μm, $f_i=(w_{1c}+w_{2c})/2w_{ic}$, $w_{1c}=0.40$ μm, and $w_{2c}$ is the control parameter. Overall, $l$ and $m$ are independent numbers within [-1, 1], modulating the propagation constant and coupling coefficient. $l$ and $m$ can form two parameter spaces and when incorporating the 1D Bloch wave vector $k$ along the transverse direction $x$, they construct a 3D synthetic-reciprocal space $(k, l, m)$. A twofold degenerate point appears at $(k_c, l_c, m_c)=(\pi/\Lambda, 0, 0)$. We define three dimensionless coefficients $\delta l=l-l_c$, $\delta m=m-m_c$, and $\delta k=(k-k_c)/k_0$ $(k_0=\pi/\Lambda)$ and the tight-binding Hamiltonian writes

$$H = \beta_1\left(\delta l\right)\sum_j a_{1,j}^{\dagger}a_{1,j} + \beta_2\left(\delta l\right)\sum_j a_{2,j}^{\dagger}a_{2,j}$$
$$+\kappa_1\left(\delta m\right)\sum_j\left(a_{1,j}^{\dagger}a_{2,j}+a_{2,j}^{\dagger}a_{1,j}\right) + \kappa_2\left(\delta m\right)\sum_j\left(a_{1,j+1}^{\dagger}a_{2,j}+a_{2,j}^{\dagger}a_{1,j+1}\right),$$

(1)

where $\beta_{1(2)}(\delta l)$ is the propagation constant of the waveguide (i.e., the mode constant of the fundamental transverse electric (TE) waveguide mode [46]). $\kappa_{1(2)}(\delta m)$ represents coupling coefficient. Upon Fourier transformation and expanding $H$ with respect to $(\delta l, \delta m, \delta k)$ up to the first order, we finally get the effective Hamiltonian [46]:

$$H=2c\delta m\sigma_x+K_0\delta k\sigma_y+b_+\delta l\ (\sigma_z+\alpha_{\mathrm{Weyl}}\sigma_0)+\beta_-\sigma_z+\beta_+\sigma_0,$$

(2)

where $\sigma_x$, $\sigma_y$, and $\sigma_z$ are Pauli matrices, $\sigma_0$ is a 2×2 identity matrix. This Hamiltonian is a standard Weyl Hamiltonian, which contains three real parameters: the momentum $\delta k$ and parameters of $\delta l$ and $\delta m$ to mimic the synthetic momenta. The coefficients $c=\partial\kappa/\partial\delta m|_{\delta m=0}$, $K_0=-\kappa_c k_0\Lambda$ ($\kappa_c$ is the coupling coefficient at the exact Weyl point), and $b_+=(b_1+b_2)/2$ are determined by the waveguide structure parameters which refer the concept of Fermi velocity in an effective Weyl Hamiltonian in the electronic context [46,48]. The tilting of the dispersion cone is determined by the Weyl parameter $\alpha_{\text{Weyl}} = b_- / b_+ = (b_1 - b_2) / (b_1 + b_2)$, with $\alpha_{\text{Weyl}} < 1 \ (>1)$ corresponding to a type-I (-II) Weyl system [27,30]. Here, $b_{1(2)} = \partial\beta_{1(2)}/\partial\delta l|_{\delta l=0} = (w_{1c} + w_{2c})/2 \ (\partial\beta_{1(2)}/\partial w_{1(2)}|_{w1(2)=w1(2)c})$ is the first derivative of the propagation constant with respect to the synthetic parameter $\delta l$, which scales with the dispersion of waveguide modes i.e., the variation of $\beta$ as a function of $w$. For conventional waveguides, $\beta$ increases as $w$ increases ($b_{1(2)}>0$), so $\alpha_{\text{Weyl}}$ is always smaller than 1, and only type-I WPs can be achieved. Here, the SWG waveguides provide new control parameters to manipulate the propagation constants [46] and it is possible to realize conventional ($b_1>0$) and anomalous ($b_2<0$) waveguide dispersions (see Fig. 1(b)), which imply type-I and –II WPs, respectively. By varying $w_{2c}$, a continuous transition (at $w_{2c}=0.43$ μm) from a type I Weyl system to a type II can be realized (see Fig. 1(c)).

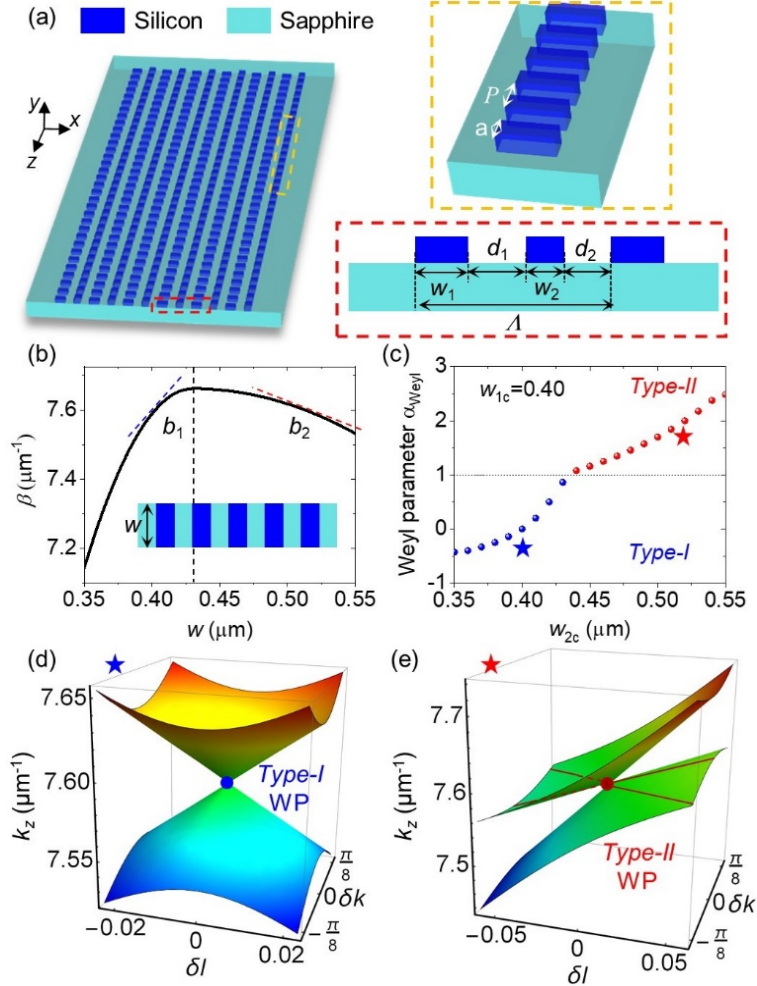

Fig. 1 (a) Synthetic Weyl lattice made by SWG waveguides. The right panels show the zoom-in structure and the cross-section of the unit cell. (b) Effective propagation constants of a single SWG waveguide. The blue and red dashed curves correspond to the positive and negative waveguide dispersions. (c) Variation of $\alpha_{\text{Weyl}}$ with $w_{2c}$. Projected band structures at the $\delta l$-$\delta k$ space ($\delta m$=0) for $w_{2c}$=0.40 μm (d) and 0.52 μm (e), representing the type-I and -II WPs (marked by the blue and red dots), respectively. The red curves in (e) show a crossing at the type-II WP.

Figures 1(d) and 1(e) show the projection of the bands in the ($\delta l$, $\delta k$) space with $\delta m$=0 for $w_{2c}$=0.4 μm (type-I) and $w_{2c}$=0.52 μm (type-II), respectively (blue and red pentagrams in Fig. 1(c)). For $w_{2c}$=0.4 μm, the two bands have opposite dispersion and their crossing forms a type-I WP with a point-like Fermi surface (Fig. 1(d)). In contrast, the dispersion is strongly anisotropic, with both bands having positive group velocities for $w_{2c}$=0.52 μm, which corresponds to a type-II WP with conical Fermi surface (Fig. 1(e)). We experimentally confirm the WPs by observing the conical diffraction effect according to their linear dispersion [46].

We connect two synthetic Weyl lattices (medium-A and -B) with independent WPs to construct the Weyl heterostructures (Fig. 2(a), left panel). Two rotational loops around the two WPs of the heterostructures in the $\delta l$-$\delta m$ parameter spaces are defined with $\delta l_{A(B)}= \delta l_0 \cos\phi_{A(B)}$ and $\delta m_{A(B)}= \delta m_0 \sin\phi_{A(B)}$, where $\phi_{A(B)} \in [0, 2\pi]$ are the rotational phases around the two WPs, $\delta l_0$ and $\delta m_0$ are the shared radii (Fig. 2(a), right panel). In this way, we can map the original two "independent" WPs to a ($\phi_A$, $\phi_B$) phase diagram. The projected band structures in the $\delta k$ space of medium-A and –B are hyperbolic curves separated by a gap (see Fig. 2(d)). For the type-I Weyl heterostructures, the gaps of two media are always matched with each other during the rotation (Fig. 2(d), right panel). However, as the important comparison for the type-II Weyl heterostructures, the two gaps shift with each other during the rotations, resulting in completely mismatched bandgaps (Fig. 2(d), middle panel) or partially overlapping bandgaps (Fig. 2(d), left panel). We define a *band-shift index* $\upsilon$ to describe the relationship between two bandgaps, where $\upsilon$=0 indicates completely mismatched bandgaps and $\upsilon$=1 represents that these two bandgaps have overlaps [46]. Figure 2(b) shows a 2D map of $\upsilon(\phi_A, \phi_B)$ for the type-II heterostructures with respect to $\phi_A$ and $\phi_B$, where the blue regions indicate totally mismatched bandgaps ($\upsilon$=0) and the red regions indicate that there are some overlaps for the two gaps ($\upsilon$=1).

Considering that the two Weyl structures rotate in opposite directions ($\phi_B$=-$\phi_A$+$\phi$, where $\phi \in [0, 2\pi]$ is a constant, Fig. 2(a), right panel). In this situation, the topological charges of the two WPs are opposite that correspond to reversed topological order [12,16,17,27] and indicates the topological modes [49,50]. For type-I Weyl heterostructures, the topological states always exist as long as the two Weyl structures rotate in opposite directions [16]. However, for the type-II Weyl heterostructures, there are some regions of $\phi$ where the bound states can become extended due to the mismatched bandgaps. To illustrate this, we define the band-shift index $\upsilon$ for different rotational loops with $\phi_B = -\phi_A + \phi$ (each $\phi$ corresponds to a rotational loop), i.e., $\upsilon(\phi) \equiv \upsilon(\phi_A = 0, \phi_B = \phi)$, as shown in Fig. 2(c). As $\phi$ increases from 0 to $2\pi$, the bound-extended-bound mode transition process is predicted, corresponding to $\upsilon(\phi)$ taking 1-0-1. The detailed condition for the bound-extended mode transition is analyzed in [46], together with the calculated inverse participation ratio (IPR). In the following, we take $\phi = 2\pi$ [$\upsilon=1$, case (i) in (b)] and $\pi$ [$\upsilon=0$, case (ii) in (b)] as two representative examples to illustrate their eigenvalue spectra and bound/extended modes.

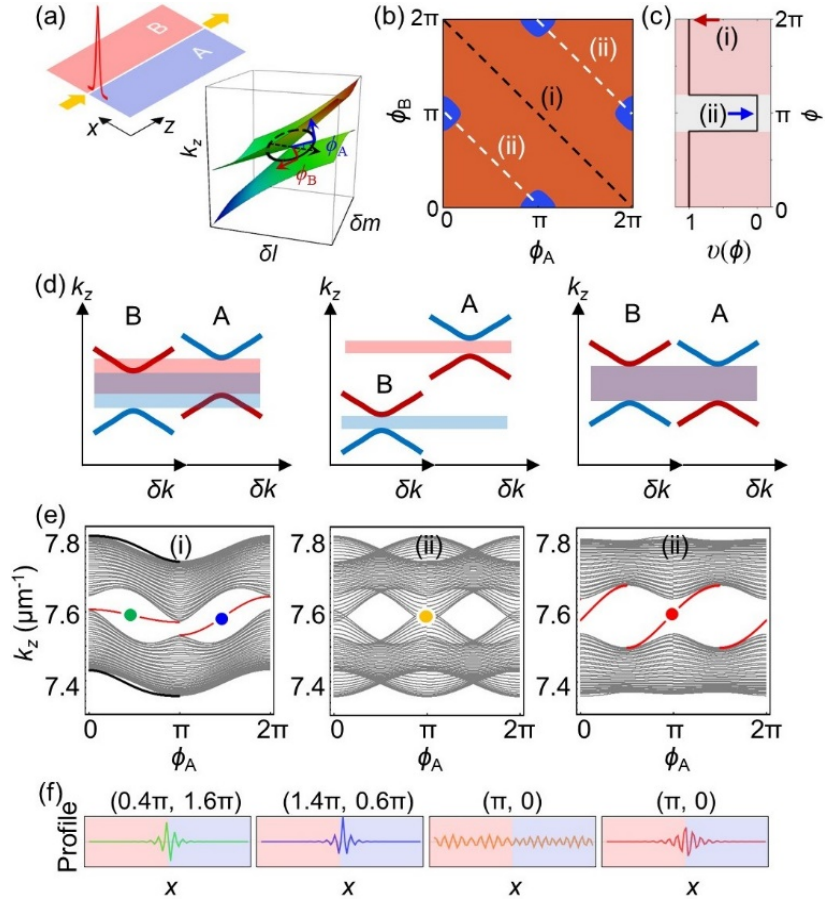

Fig. 2 (a) Constructing Weyl heterostructures with two Weyl media (A and B) (left panel). Two rotational loops ($\phi_A$ and $\phi_B$) around the WPs are introduced in $\delta l$-$\delta m$ spaces (right panel). (b) A 2D map ($\phi_A$, $\phi_B$) of $\upsilon$ shows matched (red regions, $\upsilon=1$) and mismatched (blue regions, $\upsilon=0$) gaps in type-II Weyl heterostructures. (c) Band-shift index $\upsilon$ as a function of $\phi$, where the red and blue arrows mark the case (i) and (ii) shown in (e), respectively. (d) Projected band structures in the $\delta k$ space of medium-A and –B. (e) Eigenvalue spectra as a function of $\phi_A$ for loops shown in (b), case (i) for the black line and case (ii) for the white line. The red (black) curves represent the topological bound states (trivial defect states). $\delta l_0$=0.025, $\delta m_0$=0.2, and the number of unit cells is 20 at each side. (f) Eigenmode distributions for the cases marked by dots in the spectra.

Figure 2(e) shows the eigenvalue spectra of the type-II Weyl heterostructure as a function of rotating phase $\phi_A$ with $\phi_B$=$-\phi_A$+2$\pi$ ($\phi$=2$\pi$, $\upsilon$=1) and $\phi_B$=$-\phi_A$+$\pi$ ($\phi$=$\pi$, $\upsilon$=0), corresponding to the dashed black [case (i)] and white [case (ii)] lines marked in Fig. 2(b), respectively. As expected, there are gapless topological states (red curves) connecting the upper and lower bulk bands (gray curves) for case (i) (Fig. 2(e), left panel), which exhibits antisymmetric (0<$\phi_A$<$\pi$) or symmetric ($\pi$<$\phi_A$<2$\pi$) features (see corresponding modal profile in Fig. 2(f), green curve for antisymmetric mode with ($\phi_A$, $\phi_B$) = (0.4$\pi$, 1.6$\pi$) and blue curve for symmetric one at (1.4$\pi$, 0.6$\pi$). The wavefunctions of these topological bound states can be found in [46]. To be mentioned, there are also a pair of non-topological defect modes existing beyond the bulk band (marked by black curves in Fig. 2(e)(i)), which are induced by the coupling between two media [46].

In contrast, for case (ii) crossing the blue regions ($\upsilon$=0), we find that even if the topological charges of the two WPs are still opposite, the expected topological states cannot exist (see eigenvalue spectra in Fig. 2(e), middle panel). This is because the two type-II Weyl media have totally mismatched band gaps in the blue regions. As such, only extended bulk modes can appear [see mode distributions in Fig. 2(f), orange curve, for ($\phi_A$, $\phi_B$) = ($\pi$, 0)]. Note the absence of topological modes only happens for the type-II Weyl heterostructures, while for surface modes in a uniform Weyl medium, their existence is solely dependent on the nontrivial topological charges of WP, because the gap always matches with the trivial ambient medium. In addition, we also analyze the case of type-I Weyl heterostructure for comparisons (Fig. 2(e), right panel). As expected, the interface states reemerge for case (ii) [see Fig. 2(f), red curve, ($\phi_A$, $\phi_B$) = ($\pi$, 0)].

In experiments, we fabricate the samples in a silicon wafer on a sapphire substrate [46] (Fig. 3(b)). The light (1550 nm laser) is injected into the SWG arrays via a grating coupler, a strip waveguide, and an inversed taper. The taper structure is designed with an adiabatically narrowed width to transform the strip waveguide mode to the desired SWG mode [31,32] (Fig. 3(a), top panel). Besides, for the excitation of antisymmetric bound states, two-waveguide inputs with out-of-phase ($\pi$ phase shift) are required to match the profile of eigenmodes. So we designed two inversed tapers with different variations of the widths to realize different phase accumulations and the required out-of-phase excitations (Fig. 3(a)). Figure 3(c) displays the optical propagations in experiments for the antisymmetric bound modes [Type-II Weyl heterostructures with $(\phi_A, \phi_B) = (0.4\pi, 1.6\pi)$], the light comes out from the center of the sample, indicating the emergence of the localized interface modes. The zoom-in of the output signal is shown in the right panel of Fig. 3(e), with the simulated light evolution displayed in the left. A well-trapped propagation at the interface can be observed that is consistent with the experiment. The corresponding results for the symmetric bound modes, i.e., Type-II Weyl heterostructures at $(1.4\pi, 0.6\pi)$, are also shown in Fig. 3(d). Both simulation and experiment demonstrate the localization at the interface, confirming the existence of the localized bound states. In contrast, for Type-II Weyl heterostructures at $(\pi, 0)$, the input light clearly scattered out into the bulk of the Weyl heterostructures (Fig. 3(f)), indicating no bound states at the interface. We also examine the type-I case with the same position $(\pi, 0)$ at the synthetic space (Fig. 3(g)), the localized bound mode reemerges with a strong localization at the interface.

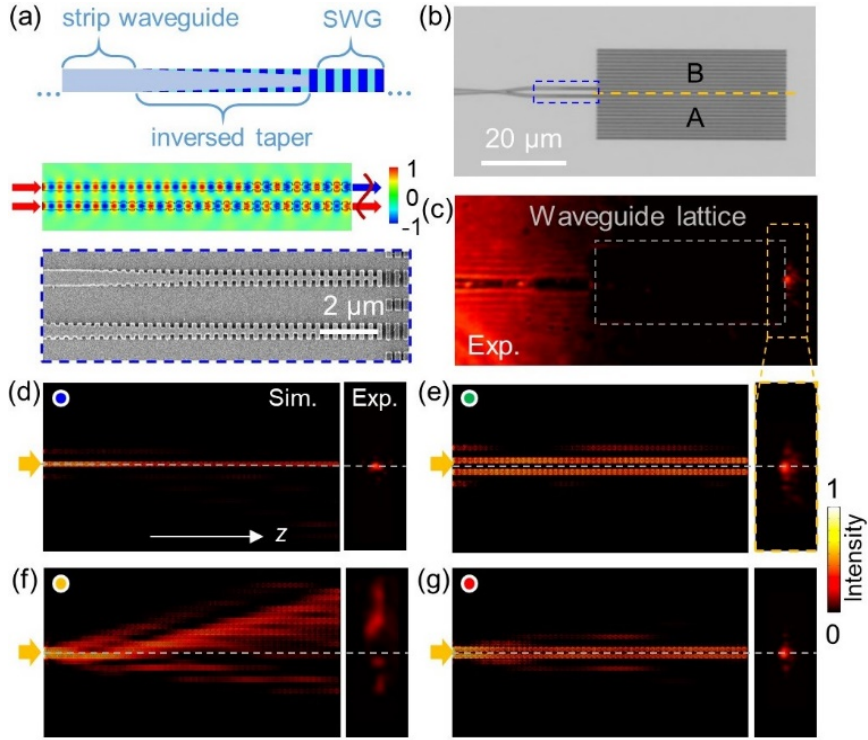

Fig. 3 (a) Top panel: Schematics of the adiabatic inversed taper. Middle panel: Simulation results of the designed pair of inversed tapers to generate two out-of-phase outputs. Bottom panel: Zoom-in top view of the fabricated inversed tapers. (b) Optical images of an experimental sample. (c)

Experimentally captured optical signals through the waveguide lattice. (d-g) Simulated light propagations (left) and experimentally detected output intensities (right) for Type-II Weyl heterostructures at $(1.4\pi, 0.6\pi)$ (d), at $(0.4\pi, 1.6\pi)$ (e), at $(\pi, 0)$ (f), and Type-I Weyl heterostructures at $(\pi, 0)$ (g), corresponding to the cases marked by blue, green, orange, and red dots in Fig. 2. The dashed lines indicate the interface.

The unique feature of the complete band mismatching for type-II Weyl heterostructures also implies quite distinct light refraction phenomena at the interface. For the normal type-I case with $\phi_A = \pi$ and $\phi_B = 0$, the projected band structures of A and B have similar parabolic dispersions with matched $k_z$ (Fig. 4(a)). Therefore, the light input from medium-B would go through the interface and enter into medium-A (Fig. 4(c)). In contrast, the projected band structures of A and B of the type-II case are shifted along $k_z$ direction due to the anisotropic dispersion, so the wave vectors on either side of type-II WP do not match each other (Fig. 4(b)). As a consequence, the input light from medium-B would be totally reflected at the interface (Fig. 4(d)). The distinguished total reflection behaviors of type-II Weyl heterostructures are well confirmed in the simulations (Fig. 4(f)), which is coincidence with the experiments (Fig. 4(f), right panel). As a comparison, the light passes through the interface and enters into medium-A for the type-I case (Fig. 4(e)). Changing the input ports or the incident angles will not influence the total reflection behaviors [46].

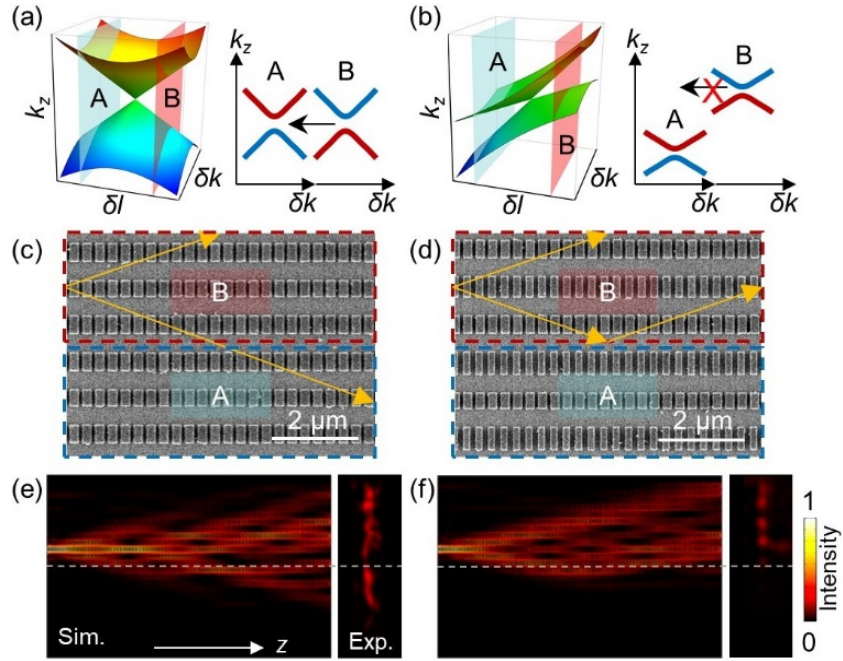

Fig. 4 (a) Left panel: Projected band structure of type-I WP in the $\delta l - \delta k$ synthetic space, the blue and red cut planes mark the positions of the medium-A and –B. Right panel: Projected bands on the two cut planes. (c) Zoom-in picture of the type-I Weyl heterostructure sample, clearly showing the SWG structures at the interface of medium-A and -B. The yellow arrows indicate the energy flow. (e) Simulated light propagations (left) and experimentally detected output intensities (right) for type-I Weyl heterostructure. (b,d,f) Corresponding results for the type-II Weyl heterostructures.

The Weyl systems and interfaces have also been studied in the context of superconductors and Andreev reflection [51-54], where the anisotropic energy spectrum of the type-II Weyl media plays a key role in tuning the Andreev reflection property. Moreover, besides the specific total reflection behaviors of the type-II case shown in Fig. 4(f), it is possible to realize partial transmission across the interface by controlling the relative values of $\phi_A$ and $\phi_B$, as well as the excitation condition, a phenomenon related to the field of Klein tunneling [55-59]. More interestingly, the Weyl semimetals and heterostructures have also generated interest in the context of analogue gravity [60-66]. Therefore, we anticipate that our experiments demonstrated in the synthetic photonic systems will attract broad interest from the community of topological physics, integrated photonics, condensed matter physics, and quantum simulations.

In summary, we explore synthetic Weyl heterostructures and demonstrate the topological manipulation of bound states and light reflection in type-II Weyl heterostructure. These Weyl media are constructed by nanostructured photonic waveguide, which gives rise to continuous controllability for building type-I and -II WPs. By connecting two independent type-II Weyl media with distinct topological charges, a type-II Weyl heterostructure can be constructed that supports topologically protected bound states, which would disappear and reemerge by tuning their rotational phases. The absence of topological mode arises from the tilted dispersion of type-II Weyl band structure. Moreover, the tilted dispersion of type-II Weyl band structure can lead to the total reflection at the interface, in contrast to the transmission for normal type-I cases. This work provides a versatile photonic platform for constructing synthetic dimensions and studying high-dimensional topological phenomena. The demonstration of Weyl heterostructure associated with novel physical consequences would inspire further exploration in a plethora of systems, ranging from photonics and microwaves to cold atoms and acoustics.

**Acknowledgments**

This research was supported by the National Key R&D Program of China (2022YFA1404301 and 2023YFA1407200) and National Natural Science Foundation of China (Nos. 12204233, 12174186, 12122407, 62288101, 92250304, and 62325504). Tao Li thanks for the support from Dengfeng Project B of Nanjing University. Luqi Yuan thanks for the sponsorship from Yangyang Development Fund.

**References**

[1]   H. Weyl, Elektron und gravitation, I. Z. Phys. 56, 330 (1929).

[2]   X. Wan, A. M. Turner, A. Vishwanath, and S. Y. Savrasov, Topological semimetal and Fermi-arc surface states in the electronic structure of pyrochlore iridates, Phys. Rev. B 83, 205101 (2011).

[3]   S.-Y. Xu, I. Belopolski, N. Alidoust, M. Neupane, G. Bian, C. Zhang, R. Sankar, G. Chang, Z. Yuan, C. Lee, S. Huang, H. Zheng, J. Ma, D. S. Sanchez, B. Wang, A. Bansil, F. Chou, P. P. Shibayev, H. Lin, S. Jia, and M. Z. Hasan, Discovery of a Weyl fermion semimetal and topological Fermi arcs, Science 349, 613 (2015).

[4]   M. Xiao, W.-J. Chen, W.-Y. He, and C. T. Chan, Synthetic gauge flux and Weyl points in acoustic systems, Nat. Phys. 11, 920 (2015).

[5]   H. He, C. Qiu, L. Ye, X. Cai, X. Fan, M. Ke, F. Zhang, and Z. Liu, Topological negative refraction of surface acoustic waves in a Weyl phononic crystal, Nature 560, 61 (2018).

[6]   F. Li, X. Huang, J. Lu, J. Ma, and Z. Liu, Weyl points and Fermi arcs in a chiral phononic crystal, Nat. Phys. 14, 30 (2018).

[7]   X. Fan, C. Qiu, Y. Shen, H. He, M. Xiao, M. Ke, and Z. Liu, Probing Weyl physics with one-dimensional sonic crystals, Phys. Rev. Lett. 122, 136802 (2019).

[8] L. Lu, Z. Wang, D. Ye, L. Ran, L. Fu, J. D. Joannopoulos, and M. Soljačić, Experimental observation of Weyl points, Science 349, 622 (2015).

[9] L. Lu, L. Fu, J. D. Joannopoulos, and M. Soljačić, Weyl points and line nodes in gyroid photonic crystals, Nat. Photon. 7, 294 (2013).

[10] W. Chen, M. Xiao, and C. T. Chan, Photonic crystals possessing multiple Weyl points and the experimental observation of robust surface states, Nat. Commun. 7, 13038 (2016).

[11] B. Yang, Q. Guo, B. Tremain, R. Liu, L. E. Barr, Q. Yan, W. Gao, H. Liu, Y. Xiang, J. Chen, C. Fang, A. Hibbins, L. Lu, and S. Zhang, Ideal Weyl points and helicoid surface states in artificial photonic crystal structures, Science 359, 1013 (2018).

[12] J. Noh. S. Huang, D. Leykam, Y. D. Chong, K. P. Chen, and M. C. Rechtsman, Experimental observation of optical Weyl points and Fermi arc-like surface states, Nat. Phys. 13, 611 (2017).

[13] M. Xiao, Q. Lin, and S. Fan, Hyperbolic Weyl point in reciprocal chiral metamaterials, Phys. Rev. Lett. 117, 057401 (2016).

[14] Q. Lin, M. Xiao, L. Yuan, and S. Fan, Photonic Weyl point in a two-dimensional resonator lattice with a synthetic frequency dimension, Nat. Commun. 7, 13731 (2016).

[15] Q. Wang, M. Xiao, H. Liu, S. Zhu, and C. T. Chan, Optical interface states protected by synthetic Weyl points, Phys. Rev. X 7, 031032 (2017).

[16] Z. Yan, Q. Wang, M. Xiao, Y. Zhao, S. Zhu, and H. Liu, Probing rotated Weyl physics on nonlinear lithium niobate-on-insulator chips, Phys. Rev. Lett. 127, 013901 (2021).

[17] W. Song , S. Wu, C. Chen, Y. Chen, C. Huang, L. Yuan, S. Zhu, and T. Li, Observation of Weyl interface states in non-Hermitian synthetic photonic systems, Phys. Rev. Lett. 130, 043803 (2023).

[18] O. Boada, A. Celi, J. I. Latorre, and M. Lewenstein, Quantum simulation of an extra dimension, Phys. Rev. Lett. 108, 133001 (2012).

[19] L. Yuan, Y. Shi, and S. Fan, Photonic gauge potential in a system with a synthetic frequency dimension, Opt. Lett. 41, 741 (2016).

[20] A. Dutt, Q. Lin, L. Yuan, M. Minkov, M. Xiao, and S. Fan, A single photonic cavity with two independent physical synthetic dimensions, Science 367, 59 (2020).

[21] M. Wimmer, H. M. Price, I. Carusotto, and U. Peschel, Experimental measurement of the Berry curvature from anomalous transport, Nat. Phys. 13, 545 (2017).

[22] E. Lustig, S. Weimann, Y. Plotnik, Y. Lumer, M. A. Bandres, A. Szameit, and M. Segev, Photonic topological insulator in synthetic dimensions, Nature 567, 356 (2019).

[23] X.-W. Luo, X. Zhou, J.-S. Xu, C.-F. Li, G.-C. Guo, C. Zhang, and Z.-W. Zhou, Synthetic-lattice enabled all-optical devices based on orbital angular momentum of light, Nat. Commun. 8, 16097 (2017).

[24] T. Ozawa and H. M. Price, Topological quantum matter in synthetic dimensions, Nat. Rev. Phys. 1, 349 (2019).

[25] D.-H.-M. Nguyen, C. Devescovi, D. X. Nguyen, H. S. Nguyen, and D. Bercioux, Fermi arc reconstruction in synthetic photonic lattice, Phys. Rev. Lett. 131, 053602 (2023).

[26] Y. Xu, F. Zhang, and C. Zhang, Structured Weyl points in spin-orbit coupled fermionic superfluids, Phys. Rev. Lett. 115, 265304 (2015).

[27] A. A. Soluyanov, D. Gresch, Z. Wang, Q.S. Wu, M. Troyer, X. Dai, and B. Andrei Bernevig, Type-II Weyl semimetals, Nature 527, 495 (2015).

[28] R. Li, B. Lv, H. Tao, J. Shi, Y. Chong, B. Zhang, and H. Chen, Ideal type-II Weyl points in topological circuits, Nat. Sci. Rev. 8, nwaa192 (2021).

[29] M. Tian, I. Velkovsky, T. Chen, F. Sun, Q. He, and B. Gadway, Observation of topological Weyl type I-II transition in synthetic mechanical lattices, arXiv:2308.07853 (2023).

[30] Y. Yang, W. Gao, L. Xia, H. Cheng, H. Jia, Y. Xiang, and S. Zhang, Spontaneous emission and resonant scattering in transition from type I to Type II photonic Weyl systems, Phys. Rev. Lett. 123, 033901 (2019).

[31] P. Cheben, R. Halir, J. H. Schmid, H. A. Atwater, and D. R. Smith, Subwavelength integrated photonics, Nature 560, 565 (2018).

[32] Y. Meng, Y. Chen, L. Lu, Y. Ding, A. Cusano, J. A. Fan, Q. Hu, K. Wang, Z. Xie, Z. Liu, Y. Yang, Q. Liu, M. Gong, Q. Xiao, S. Sun, M. Zhang, X. Yuan, and X. Ni, Optical meta-waveguides for integrated photonics and beyond, Light Sci. Appl. 10, 235 (2021).

[33] A. B. Redondo, I. Andonegui, M. J. Collins, G. Harari, Y. Lumer, M. C. Rechtsman, B. J. Eggleton, and M. Segev, Topological optical waveguiding in silicon and the transition between topological and trivial defect states, Phys. Rev. Lett. 116, 163901 (2016).

[34] A. B. Redondo, B. Bell, D. Oren, B. J. Eggleton, and M. Segev, Topological protection of biphoton states, Science 362, 568 (2018).

[35] W. Song, W. Sun, C. Chen, Q. Song, S. Xiao, S. Zhu, and T. Li, Breakup and recovery of topological zero modes in finite non-Hermitian optical lattices, Phys. Rev. Lett. 123, 165701 (2019).

[36] X. T. He, E. T. Liang, J. J. Yuan, H. Y. Qiu, X. D. Chen, F. L. Zhao, and J. W. Dong, A silicon-on-insulator slab for topological valley transport, Nat. Commun. 10, 872 (2019).

[37] M. I. Shalaev, W. Walasik, A. Tsukernik, Y. Xu, and N. M. Litchinitser, Robust topologically

protected transport in photonic crystals at telecommunication wavelengths, Nat. Nanotech. 14, 31 (2019).

[38] W. Song, W. Sun, C. Chen, Q. Song, S. Xiao, S. Zhu, and T. Li, Robust and broadband optical coupling by topological waveguide arrays, Laser Photon. Rev. 14, 1900193 (2020).

[39] A. Vakulenko, S. Kiriushechkina, D. Smirnova, S. Guddala, F. Komissarenko, A. Alù, M. Allen, J. Allen, and A. B. Khanikaev, Adiabatic topological photonic interfaces, Nat. Commun. 14, 4629 (2023).

[40] D. Smirnova, S. Kruk, D. Leykam, E. Melik-Gaykazyan, D.-Y. Choi, and Y. Kivshar, Third-harmonic generation in photonic topological metasurfaces, Phys. Rev. Lett. 123, 103901 (2019).

[41] H. Wang, G. Tang, Y. He, Z. Wang, X. Li, L. Sun, Y. Zhang, L. Yuan, J. Dong, and Y. Su, Ultracompact topological photonic switch based on valley-vortex-enhanced high-efficiency phase shift, Light Sci. Appl. 11, 292 (2022).

[42] M. S. Kirsch, Y. Zhang, M. Kremer, L. J. Maczewsky, S. K. Ivanov, Y. V. Kartashov, L. Torner, D. Bauer, A. Szameit, and M. Heinrich, Nonlinear second-order photonic topological insulators, Nat. Phys. 17, 995 (2021).

[43] S. Xia, D. Kaltsas, D. Song, I. Komis, J. Xu, A. Szameit, H. Buljan, K. G Makris, and Z. Chen, Nonlinear tuning of PT symmetry and non-Hermitian topological states, Science 372, 72 (2021).

[44] H. Zhao, X. Qiao, T. Wu, B. Midya, S. Longhi, and L. Feng, Non-Hermitian topological light steering, Science 365, 1163 (2019).

[45] X.-L. Zhang, F. Yu, Z.-G. Chen, Z.-N. Tian, Q.-D. Chen, H.-B. Sun, and G. Ma, Non-Abelian braiding on photonic chips, Nat. Photon. 16, 390 (2022).

[46] See Supplemental Material at XXX for more details on the theory and experiment, which includes Refs. [12,13,16,17,27,30,31,47].

[47] J. T. Edwards and D. J. Thouless, Numerical studies of localization in disordered systems, J. Phys. C: Solid State Phys. 5, 807 (1972).

[48] C. Guo, V. S. Asadchy, B. Zhao, and S. Fan, Light control with Weyl semimetals, eLight 3, 2 (2023).

[49] M. Z. Hasan and C. L. Kane, Colloquium: topological insulators, Rev. Mod. Phys. 82, 3045 (2010).

[50] B. A. Bernevig and T. L. Hughes, Topological Insulators and Topological Superconductors (Princeton University Press, 2013).

[51] Z. Hou and Q.-F. Sun, Double Andreev reflections in type-II Weyl semimetal-superconductor junctions, Phys. Rev. B 96, 155305 (2007).

[52] X.-S. Li, S.-F. Zhang, X.-R. Sun, and W.-J. Gong, Double Andreev reflections and double electron transmissions in a normal-superconductor-normal junction based on type-II Weyl semimetal New J. Phys. 20 103005 (2018).

[53] A. Azizi and B. Abdollahipour, Andreev reflection at the interface of a normal Weyl semimetal and a superconducting type-II Weyl semimetal, Phys. Rev. B 102, 024512 (2020).

[54] A. Azizi and B. Abdollahipour, Crossed Andreev reflection in FSF Weyl semimetal junctions, J. Phys. D: Appl. Phys. 55, 065302 (2022).

[55] O. Klein, Die Reflexion von Elektronen an einem Potentialsprung nach der relativistischen Dynamik von Dirac, Z. Phys. 53, 157 (1929).

[56] T. E. O'Brien, M. Diez, and C. W. J. Beenakker, Magnetic Breakdown and Klein Tunneling in a Type-II Weyl Semimetal, Phys. Rev. Lett. 116, 236401 (2016).

[57] S. Longhi, Klein tunneling in binary photonic superlattices, Phys. Rev. B 81, 075102 (2010).

[58] C. W. J. Beenakker, Colloquium: Andreev reflection and Klein tunneling in graphene, Rev. Mod. Phys. 80, 1337 (2008).

[59] D. Yu, G. Li, M. Xiao, D.-W. Wang, Y. Wan, L. Yuan, and X. Chen, Simulating graphene dynamics in synthetic space with photonic rings, Commun. Phys. 4, 219 (2021).

[60] G. E. Volovik and K. Zhang, Lifshitz Transitions, Type-II Dirac and Weyl Fermions, Event Horizon and All That, J. Low Temp. Phys. 189, 276 (2017).

[61] C. D. Beule1, S. Groenendijk1, T. Meng and T. L. Schmidt, Artificial event horizons in Weyl semimetal heterostructures and their non-equilibrium signatures, SciPost Phys. 11, 095 (2021).

[62] G. E. Volovik, Type-II Weyl Semimetal versus Gravastar, JETP Lett. 114, 236 (2021).

[63] V. Könye, C. Morice, D. Chernyavsky, A. G. Moghaddam, J. van den Brink, and J. van Wezel, Horizon physics of quasi-one-dimensional tilted Weyl cones on a lattice, Phys. Rev. Res. 4, 033237 (2022).

[64] D. Sabsovich, P. Wunderlich, V. Fleurov, D. I. Pikulin, R. Ilan, and T. Meng, Hawking fragmentation and Hawking attenuation in Weyl semimetals, Phys. Rev. Res. 4, 013055 (2022).

[65] V. Könye, L. Mertens, C. Morice, D. Chernyavsky, A. G. Moghaddam, J. van Wezel, and J. van den Brink, Anisotropic optics and gravitational lensing of tilted Weyl fermions, Phys. Rev. B 107, L201406 (2023).

[66] A. Haller, S. Hegde, C. Xu, C. De Beule, T. L. Schmidt, and T. Meng, Black hole mirages: electron lensing and Berry curvature effects in inhomogeneously tilted Weyl semimetals, SciPost Phys. 14, 119 (2023).